# Multiple Presents: How Search Engines Re-write the Past

*New Media & Society* (forthcoming)

Iina Hellsten,[i] Loet Leydesdorff,[ii] and Paul Wouters[iii]

***Abstract***

Internet search engines function in a present which changes continuously. The search engines update their indices regularly, overwriting Web pages with newer ones, adding new pages to the index, and losing older ones. Some search engines can be used to search for information at the internet for specific periods of time. However, these 'date stamps' are not determined by the first occurrence of the pages in the Web, but by the last date at which a page was updated or a new page was added, and the search engine's crawler updated this change in the database. This has major implications for the use of search engines in scholarly research as well as theoretical implications for the conceptions of time and temporality. We examine the interplay between the different updating frequencies by using *AltaVista* and *Google* for searches at different moments of time. Both the retrieval of the results and the structure of the retrieved information erodes over time.

**Keywords:** search engines, internet, time, temporality

[i] The Virtual Knowledge Studio for the Humanities and Social Sciences, Royal Netherlands Academy of Arts and Sciences, P.O. Box 95 110, 1090HC Amsterdam, The Netherlands; e-mail: iina.hellsten@vks.knaw.nl ; www.virtualknowledgestudio.nl

[ii] University of Amsterdam, Amsterdam School of Communications Research (ASCoR), Kloveniersburgwal 48, 1012 CX Amsterdam, The Netherlands; e-mail: loet@leydesdorff.net; www.leydesdorff.net

[iii] The Virtual Knowledge Studio for the Humanities and Social Sciences, Royal Netherlands Academy of Arts and Sciences, P.O. Box 95 110, 1090 HC Amsterdam, The Netherlands; e-mail: Paul.Wouters@vks.knaw.nl; www.virtualknowledgestudio.nl

## *Introduction*

Web pages in the internet are updated with varying frequencies. Archived Web pages, such as citation index databases, on-line archives, and postings in discussion groups remain usually static over time. Newspaper headlines, at the other end of the spectrum, are sometimes updated even hourly, and in between there is a wide scale of updating frequencies. The discrepancy between 'static' and 'dynamic' Web pages has not been studied in detail in internet research or communication studies nor have there been studies in these fields of how this affects the study of the internet. As we will explain in more detail later in this article, search engines generate a particular user experience of 'the present' in the Web, by generating links to information that seems to be presently available at the time of the search. Because each search engine generates a present every time a user enters a search query, we suggest to consider the result as multiple of presents. Our aim is to study how this constantly changing definition of the present affects the use of search engines for research purposes in the social sciences and humanities. We approach this question by empirically studying the changing presents of the Internet search engines results.

Search engines have been studied from the point of view of the currency of the information in their database indexes (Brewington & Cybenko, 2000), instabilities in the results (Bar-Ilan, 1999, 2001; Bar-Ilan & Peritz, 2002), economical and language-based inequalities in the search engine results (Introna & Nissenbaum, 2000; Vaughan & Thelwall, 2004; Van Couvering, 2004), and the lack of interactivity on the Web (Wouters & Gerbec, 2003). Most studies focus on the performance of various search engines from the point of view of a general user (Risvik & Michelsen, 2002; Lewandowski, 2004). Our focus is not on general users nor on search engine performance but on the theoretical and practical implications of search engine use for scholarly research. The way search engines re-write the past by updating their indexes in the present has hitherto received little attention (Wouters *et al*., 2004). In this paper, we address a set of questions relating to how search engines can be considered as 'clocks' of the internet that tick with different frequencies. More specifically, we are interested in the way the updating affects the present that is produced by search engines and in which they evolve.

The question of how temporal representations change over time is an urgent one. In every social reality, temporality is central to the network of relationships. Societies reconstruct themselves by reconstructing also their histories. This can be considered as a constant process of mutual adaptation between historical traditions and institutions, and between emerging expectations about the future and appreciations of the past (Schütz, 1932). The duration of activities and processes, and the ways in which they are synchronized and updated, affect the positions of agents in the network. The network development itself can be considered as an interplay and interaction effect among the various temporalities involved (Innis, 1952; Nowotny, 1994).

In terms of systems theory, this can be understood as an interference among the updating frequencies of the subsystems in society. The subsystem of science, for example, publishes scientific results with a frequency very different from that of newspapers. Similarly, some Web pages are updated with a frequency higher than others, and different search engines update their indexes with structurally different frequencies (Thelwall, 2001). Furthermore, new pages are continuously added to the Web and old ones are removed from the Web. A focus on the different updating frequencies and their temporality enables the analysis of socio-technical systems in which technical constructs are functioning both as nodes and as media facilitating relationships between the nodes of the network (Latour, 1988; Leydesdorff, 1994, 2001).

The study of updating cycles has an especially salient relevance to search engines. Some search engines (for example, *AltaVista* and *Google*) can be used to search for information in the internet for specific periods of time.[1] However, these 'date stamps' are not determined by the first occurrence of the pages in the Web, but by the last date at which a page was updated or a new page was added and the search engine's crawler updated this change in the database. For the update in the search engine database, any alteration of the Web page may count as a change, no matter how minor it was. The

[1] Note that other search engines, such as AlltheWeb, also provide the option for time limited searches but only in the form of 'past 6 months' or 'past year' while *AltaVista* and *Google* provide the option for limiting the searches to specific dates in the form of dd/mm/yy, from 01/01/02 to 31/12/02 for example.

'same' Web page may therefore belong to the year 1995 in a data set collected in 2003, while in a data set collected in 2004 it belongs to the year 2003—or it may have been 'forgotten' by the search engine altogether (Bar-Ilan, 1999). Hence, when they are used to search for historical dates, search engines represent the *results* of the interacting frequencies of (a) the creation and updating of Web pages and (b) the retrieval and updating at the level of search engine indices. The results are not likely to reflect the dates of publication of the documents under study. This has implications for the use of search engines in scholarly research.[2]

While the development of the engines remains historical, their dynamics evolve in the present and reflexively to the system to which they belong (that is, the internet). Thus, these engines reconstruct their histories by looking backwards. In other words, search engines provide the past with a 'meaning' and can thus be considered as anticipatory systems (Rosen, 1985; Dubois, 1998; Leydesdorff, 2005). Because of the updating effects, such reconstructions will tend to draw Web sites into the most recent past, thereby possibly erasing the older representations of the same Web pages. Search engines catalogue the Web, and these catalogues are continuously updated in order to keep them current.

## *Research Questions*

In this study we attempt to test how the three updating frequencies (updating the Web pages, updating the search engine database, and the growth of the Web) resonate at the internet. Search engine results allow us to study empirically the constant change in the multiple presents. We compared two search engines by performing searches with exactly the same search string at different moments of time. The focus was on the two major search engines that provide the option to limit searches to specific dates. *AltaVista's Advanced Search Engine* (www.AltaVista.com/web/adv) allows searches from the year 1980 to the present, limited to specific dates, months, or years. *Google* is currently the most frequently used and largest search engine

[2] Internet Archive (www.archive.org) aims at archiving Web pages for historical analyses of the Web, but currently it is neither complete in particular domains nor representative of parts of the Web, and it lacks the option for key word based searches in the archive.

(www.searchengineshowdown.com). It provides the option for similar date-specific searches via *Google*'s APIs or *Faganfinder* (www.faganfinder.com/Google.html). The latter engine exploits the database of *Google*.[3]

Originally, we planned to provide search results with a one-year time interval (January 2003 versus January 2004) and a one-month time-interval (January 2004 and February 2004). During our study, however, *AltaVista* changed its search engine to the one of *Yahoo!* (April 2004). The number of hits thereafter declined considerably, and therefore we decided to conduct an additional search at the end of April 2004. In general, search engines function very differently. The exact algorithms used by the various engines are commercial secrets, but it is known that while *Google* uses link-based crawling for updating its database, *Altavista* relies on a keyword-based crawling (www.searchenginewatch.com).

We are interested in two related questions. One is the question of the extent to which the same results can be reproduced using search engines for searches at different moments of time, i.e. at different 'presents'. Because of the updating mechanisms, one can no longer assume that time-series data reflect historical developments of the systems under study. This raises the question whether one can construct time series data by periodically searching the Web for specific retrieval terms. To which extent can these results be reproduced? What does the level of reproducibility reveal about the resonance between the various updating frequencies?

The second question is related: How can the changes in the results be interpreted? It seems too easy to conclude that this type of data is worthless, since the 'errors' are generated systematically. The updating mechanism represents a significant socio-technical activity on the Web. At the same time, the updating of the Web pages provides us with an empirical domain to study this mechanism of change. What kind of windows on the reality of the Web do the search engines provide?

Before addressing the technical details of the experiment, let us first specify our theoretical expectations with reference to the debate about the nature of time in these digital networks. Thereafter, we explain our experiment and its results. The last section is devoted to the methodological and substantive conclusions.

[3] Google uses the Julian calendar, but the FaganFinder automatically converts calendar dates into this older time scale.

### *Time and the internet*

In many different ways, the internet has conveyed the notion that it somehow has a profound effect on the relations between space and time. The early champions of the Net were convinced of the breakdown of temporal and spatial differences by going online (Brand, 1987). Notions such as 'timeless time' (Castells, 1996, p. 464), 'simultaneity of non-simultaneous' (Brose, 2004; Laguerre, 2004), 'ultra-present' (Goldhaber, 2004) and 'extended present' (Nowotny, 1994, p. 11) all aim at characterizing the changes in our conceptions of time and temporality due to new ICTs and digital networks.

Hassan (2003) has proposed the notion of 'network time': Network time is digitally compressed clock-time, and as such operates on a spectrum of technologically possible levels of compression. This spectrum is 'open ended' (Hassan, 2003, p. 233). According to Hassan, the observed acceleration of time follows from the premise of asynchronicity among the networks, i.e., different frequencies of change: 'The "revolution" in information technologies has been to take this to another level of temporality, to compress the meter of the clock and to accelerate the time standard of modernity. The creation of the network has simultaneously created a digital environment, an information ecology that generates its own temporality' (Hassan, 2003, p. 233).

From this perspective the search engines can be considered as subsystems of the e-society which function as clocks of the internet that 'tick' at different frequencies. The search engines update their catalogues at different frequencies, and as a consequence time is reconstructed as a resonance effect between these different frequencies. Whereas modern 'clock time' was designed to gather people at one place at the same time, the internet would allow for simultaneous access to information free from physical locations, thus leading to the 'simultaneity of the non-simultaneous' (Brose, 2004; Laguerre, 2004). However, there are two opposing views on how global networks affect the interplay between time and space.

One side claims that global networks lead to the dissolution of both time and space as relevant categories, because everything can take place at the same time and largely independently of geographical constraints. From this perspective, place is no longer relevant in cyberspace. A more nuanced version of this position has been taken by Castells (1996), who claimed that the measurable clock-time of the industrial revolution is being shattered 'in the network society, in a movement of extra-ordinary historical significance.' He captured this in the concept of 'timeless time' (Castells, 1996, p. 464): 'I propose the idea that timeless time, as I label the dominant temporality of our society, occurs when the characteristics of a given context, namely, the informational paradigm and the network society, induce systemic perturbation in the sequential order of phenomena performed in that context.' Brose (2004, pp. 16-17) argues that the impression of an acceleration of time may be a result of the simultaneity of non-simultaneous, multiple presents.

A second perspective claims that the modernist clock-time, far from being dissolved, actually extends its domination through ICT and the global networks. These scholars build on the analysis of the role of technical time standardization in the rise of capitalism and more specifically the industrial revolution (Thompson, 1967). From this perspective, the central role of time has been the coordination (in the sense of control and connecting) of social relationships (Elias, 1992). The new digital technologies would play the same role, building on the social process of standardization of time made possible by the mechanical clock (Adam, 2004). Urry (2000), for example, draws a parallel between the emergence of the internet and the railway system in the $19^{th}$ century.

Telegraphy first made it possible to construct networks spanning the globe. Using international standard time (GMT) these systems could be globalized (e.g., for the purpose of air traffic control). These networks preceding the internet would already have extended the domination of standard time to parts of the world that hitherto had been relatively unaffected (Nowotny, 1994). Far from freeing individuals or groups from the regime of the clock, the internet can be expected to subsume all remaining variety to a new regime that is even stricter. This technical standardization of time would leave no room for the post-modern deconstruction of time (Adam, 2004).

In an exposé on the technicity of time, Mackenzie (2001) proposed to conceptualize clock-time as a 'temporal and topological ordering that continues to unfold from a metastability.' Mackenzie compares time measurement to the sudden crystallization in a supersaturated solution that makes the solution metastable. Metastability refers to the tension in the synchronization of different 'clocks', and multiple presents. By using this concept of metastability, Mackenzie (ibid.) wishes to combine three analytical perspectives on time: Heidegger's exteriorization of temporality, Elias's notion of the transitions between different social timing regimes, and Latour's view of the technical mediation of time. The two mechanisms of processing in a forward mode and rewriting with hindsight can also be distinguished in terms of the possibilities to stabilize or globalize a metastability (Leydesdorff, 2001).

The dominance of linear time was fueled by the industrial revolution, which enabled people to transform time into money and place a premium on the rationalization of time. [4] Like the social construction of time, however, every conception of time should take into account both its linear and cyclical dimensions. The present re-conceptualization of time builds upon the standardized world time of the industrial revolution, yet fundamentally alters it by adding cycles as older notions of time. This reconceptualization is driven by the new information and communication technologies as socio-technical practices. These technologies generate a drive for 'a world-wide condition of simultaneity' (Nowotny, 1994, p. 9). Because of the illusion that temporal and spatial differences matter less, time and space seem to be compressed and collapsed in the world of the internet into terms of globalized communications.

In summary, the concept of a single time axis which is moving forward like an arrow is broken in the post-modern appreciation of a variety of time horizons in different social systems and for the different actors involved (Coveney & Highfield, 1990; Prigogine & Stengers, 1988). Different updating and growth frequencies may resonate historically into stability (e.g., institutions), and subsequently the metastability of the resulting system can also be globalized into an order of expectations operating in the present (Husserl, 1929; Luhmann, 2002). The present is not only the fleeting, uncapturable moment between past and future, but also a broad horizon of experiences in which pasts and futures are being recycled.

[4] The linearity of time is still dominant in metaphors of time as a forward movement in space, such as 'life is a journey' or 'scientific progress' (Hellsten, 2002).

With an inspiration very similar to that of Brose's (2004) 'simultaneity of the non-simultaneous,' Goldhaber (2004) describes the mentality of the *Homo Interneticus* as being captured in an 'ultra-present' where things constantly happen. The ultra-present is not only a redefinition of the *durée* of the present, but also of a balance between linearity and cyclicality. A comparable notion is captured by Nowotny's concept of the 'extended present:' 'The permeability of the time-boundary between present and future is increased by technologies which facilitate temporal uncoupling and decentralization, and which produce different models of time referring to the present that have largely become detached from linearity' (Nowotny, 1994, p. 11). In short, the present can be considered as both the generator and the result of interacting cycles that have their own specific frequencies. The present of the search engines is created by the three updating frequencies of the Web pages, the search engine databases, and the overall evolution of the Web.

Perhaps, the internet can be seen as the embodiment of an extended present, turned from really virtual to virtually real thanks to the new technologies of virtualization (Latour, 1991). If this were the case, we should add the notion of fragmentation to that of the extended present because any resolution would necessarily remain historical. In general, the reflexive operation contains a reference to the historical situation, but that situation is looked at from the perspective of the present, i.e., with hindsight. What is precisely added by the reflexive (albeit automated) mechanism of rewriting the system (the internet) by a subsystem (search engine) of the same system? Does the feedback arrow affect the feedforward one, and if so, how? Perhaps we should amend the 'extended present' proposed by Nowotny (1994), and turn it into a notion of many competing and fragmented, multiple extended presents—in the plural? The multiple extended presents are a result of the resonances between the different updating cycles, and this can be studied empirically by the analysis of search engine results. We aim to study how this 'present' changes over time and across search engines.

### *Research Design*

Our experiments focus on how two major search engines, *AltaVista* and *Google,* have reconstructed the Web pages on ‘frankenfoods’ over time. The metaphor of ‘frankenfoods’ has been used on the Web in the debate on genetically modified foods since the mid-1990s in the pages of various consumer and environmental organizations, in discussion forums and newsletters as well as in political arguments and journalistic accounts of the debate. In these ‘static’, i.e. archived Web pages, the use of the metaphor on the Web reached its peak between 1998 and 2000, and thereafter its use decreased rapidly (Hellsten, 2003). In this study, we can contrast this result with that of ‘dynamic’, i.e. faster changing Web pages as represented in the search engine results. In other words, this search term provides us with a well delineated topic and a relatively unambiguous search term with a clear life cycle. It is interesting to see how the updating mechanisms work on a topic on which new Web pages are not likely to have been added since 2000, while the Web continues to grow all the time.

The data was initially collected on 21-23 January 2003 using only the *AltaVista Advanced Search Engine*. The searches were at that time limited to the years 1995-2002. This data collection was repeated exactly after one year, i.e., on 21-23 January 2004, and then after one month, i.e., on 21-23 February 2004, and after three months, i.e., on 21-23 April 2004. The searches in 2004 used both *AltaVista* and *Google*, and included the year 2003. The results for the year 2003 were further decomposed into the twelve months of that year in order to distinguish between the long-term and short-term effects of the updating in the different presents in more detail.

The user interfaces of the two search engines provide different options for using search terms. With *AltaVista* we originally used the search string *frankenfood* OR (frankenstein AND food*)*[5] for the retrieval. We used the *FaganFinder* interface to *Google* that allows us to use the date range capability of *Google*. However, this interface does not allow the combination of Boolean operators, and the * wildcard does not function in the 'exact phrase' option. For this reason, the original search string was split into three versions, for each of which the results were collected separately and then pooled: *frankenstein food*, *frankenstein foods* and *frankenfood(s)*.[6] In order to compare the results of *Google* with *AltaVista*, we also used the following string in *AltaVista*: *frankenstein food* OR frankenfood** for the three searches conducted in 2004.

[5] After April, 2004 the *AltaVista* no longer allows for wild cards. This, however, does not affect our study.

We not only checked the reported number of hits of each search engine, but also downloaded the pages with the search results. These pages contain the titles, first sentences, document types, and URLs of the hits. This material allows us to check how many of the reported results could actually be retrieved from the internet. More importantly, the titles provide us with a semantic domain that can be mapped and visualized in order to see how the words used in the titles of the results are positioned, and whether the clusters of words change from one data collection to another. We use techniques that were developed for this purpose in other contexts (Leydesdorff, 2004: Leydesdorff & Hellsten, 2005) and provide the visualizations below in order to illustrate our arguments with substantive interpretations.[7]

Our expectation about the changes of the different presents generated by the search engines can be formulated as follows. First, we expect that the distribution of the reported number of hits over the years will show a strong *bias in favour of the most recent year* (relative to the date of the measurement, i.e. the 'present' when the data was collected). We call this the long-term memory of search engines. Second, if it is true that Web sites are continuously overwritten with newer date stamps, then we would expect a *decrease* in the total number of hits for the months before the most recent one (again relative to the date of the measurement). We call this the short-term memory.

[6] We also tested the string *frankenstein AND food* in *Google*, but this generated many pages about food with Frankenstein movies in relation to the number of pages about the debate on genetically modified food.

[7] The mappings are based on using the so-called vector-space-model for the analysis (Salton & McGill, 1983). The program is freely available at http://www.leydesdorff.net/software/fulltext. Pajek is used for the visualizations. Pajek is freely available at http://vlado.fmf.uni-lj.si/pub/networks/pajek/ .

In addition, we are interested in the substance of the search results, i.e., the structure of the information retrieved using the search engines. We tested if the structure in the data changes. In summary, we study the construction of time both in terms of changes in the reported numbers of results per year and the actually retrieved results. We use the reported numbers for the study on short- and the long-term memory, while the semantic maps are based on the retrieved results.

## *Re-Writing the Past*

### *Long-term memory*

The long-term memory of the search engines can be expected to show a bias towards the latest year. Figures 1 and 2 show the development of the long-term memory results of *AltaVista* and *Google*, collected in January, February, and April 2004, respectively, and with exactly the same strings (*frankenfood* OR frankenstein food**).

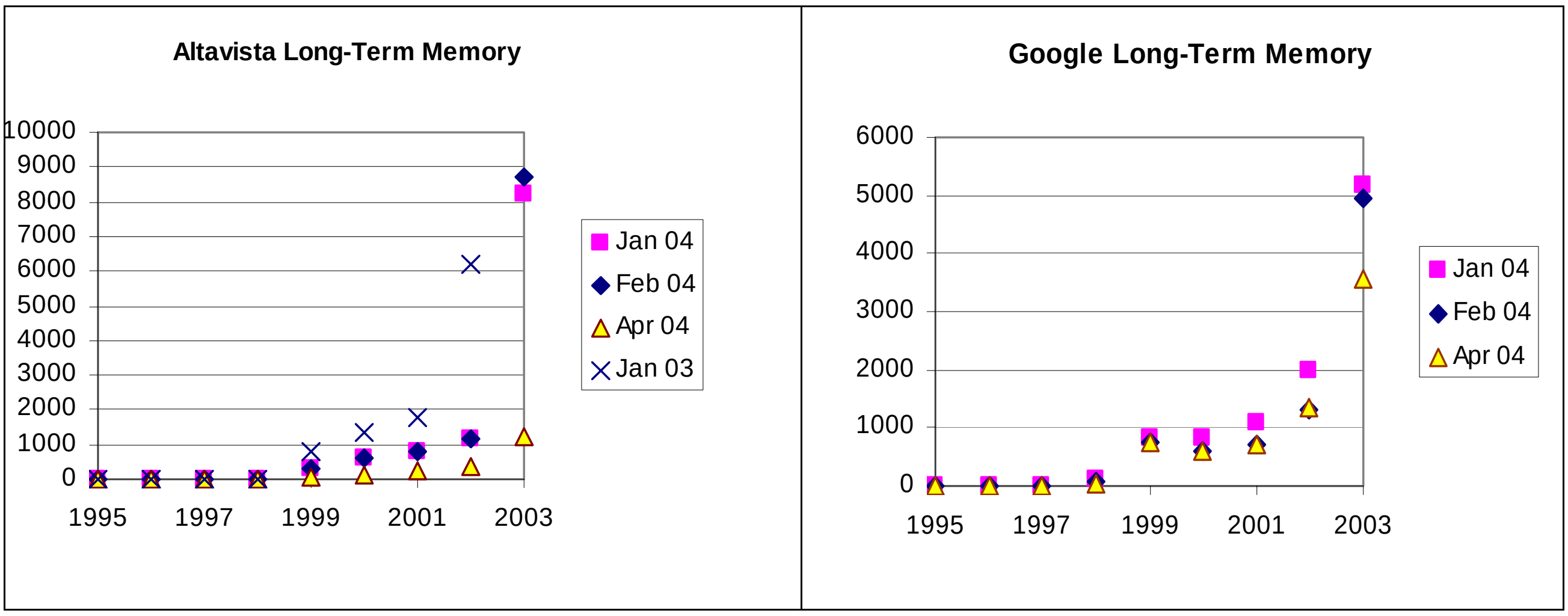

**Figures 1 and 2**: Search results using 'frankenfood* OR frankenstein food*' as search terms in Google and AltaVista.

First, the Figures 1 and 2 show that both *AltaVista* and *Google* have a strong and consistent bias towards the latest year. For the measurement in January 2003 (including only *AltaVista*), the year 2002 was the most recent year, hence this distribution is shifted one year to the left (Figure 1). The data also prove that *AltaVista* and *Google* both overwrite their histories. The sharpest fall is seen in the year 2002, which can be attributed to a massive updating of Web pages from that year in the year 2003, i.e., towards the closest to the present in question. The number of hits retrieved using *AltaVista* decreased by an order of magnitude after the search engine of *Yahoo!* was adopted. *Google* hits decrease from the measurements in January to those in April 2004 for all years. Except for the most recent year, the numbers in the measurements in February and April are at the same level.

A decrease in the overall number of results was expected because of our focus on the metaphor of 'frankenfood.' As noted, the use of this metaphor reached a peak between the years 1998 and 2000 in 'static' Web pages, and no significant number of new Web pages was expected. By using the search engines, one is able to detect also the changes in dynamic Web pages. The search engine results are remarkable since they indicate that even if the use of the metaphor has decreased, the Web pages are still updated continuously. This is consistent with the results by Bar-Ilan (1999), who found that as a result of this continuous updating, the search string is not always necessarily a part of the Web page.[8] The search engine seems to 'remember' that a certain URL was part of its database index even if the search term may have disappeared from the page.

### *Short-term memory*

According to our hypothesis, the search engines are expected to show a decrease in the total number of hits towards the most recent months relative to the month of measurement, i.e., the date of the data collection. Figures 3 and 4 show this short-term memory of the search engines.

[8] Due to the large amount of results, we were not able to check whether the search string occurs in the results we collected.

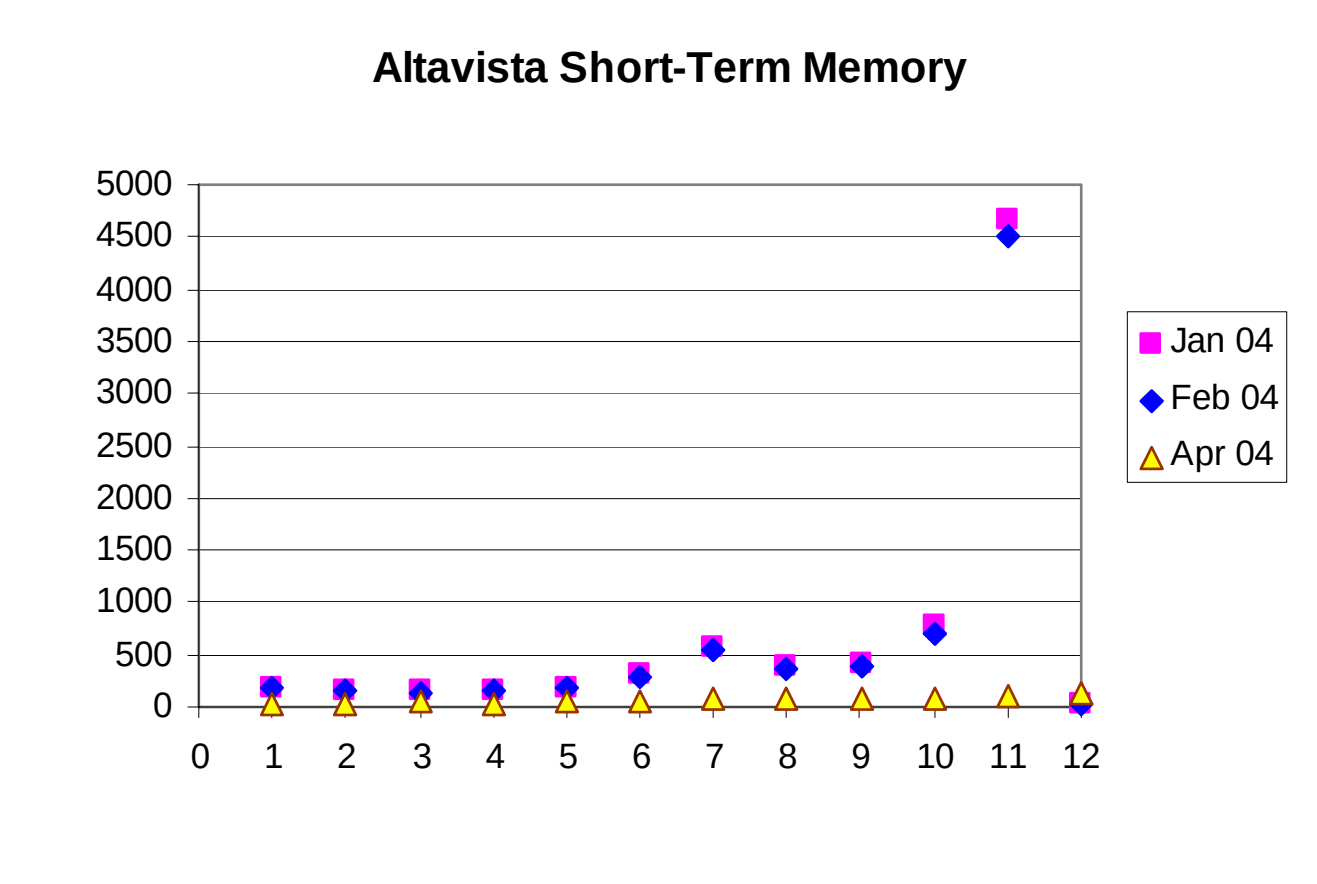

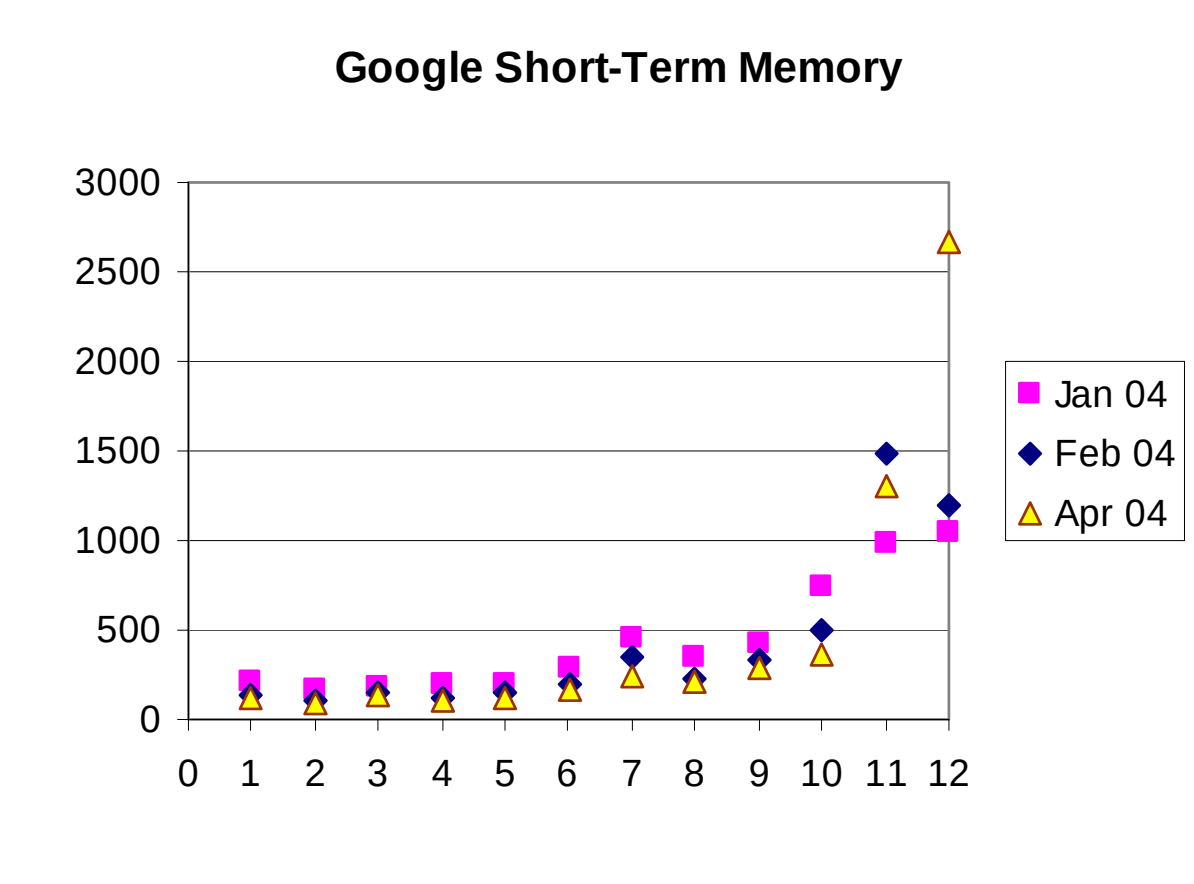

**Figures 3 and 4:** Search results using 'frankenfood* OR Frankenstein food*' as search terms in *Google* and *AltaVista*.

In the interpretation of these graphs, one should take into account that our measurements were conducted in the beginning of the year 2004 (January, February, and April). We would therefore expect a bias in favour of the months December and November 2003 if the search engines' updating frequency is high enough to be noticeable within a month. Both search engines do indeed record high numbers in the last months of the previous year, but not with the same update frequency.

The *Google* data indicate a shift over time towards the month of December. The *AltaVista* measurements in January and February 2004 are at the same level, whereas the numbers in the April measurement have fallen dramatically. We attribute this latter decline to the take-over of *AltaVista* by *Yahoo!* The *Google* data, however, substantiate the hypothesis that the historical record is being erased in the short term: the numbers for the first ten months of the year consistently decrease over time. *AltaVista* data does not show this effect. Apparently, the two search engines differ in terms of the lags and speeds in the updating of their databases.

### *Substantive similarities and differences*

We also expected that the structure of the Web pages would differ across the searches at different points of time. When the documents are relocated to more recent years, one could expect that the existing structure of documents within the year would be disrupted. Since this relocation will not be uniform for all documents, a shift of the structure as a whole into a more recent year is unlikely. Our expectation is therefore that the structure, as represented by co-appearances in the title words of the retrieved Web pages in a particular year, erodes over time. As a result, the information in the structure of the results is gradually lost.

To study this, we first calculated correlation coefficients, and then drew semantic maps based on the co-occurrences of title words of the retrieved documents. The semantic maps are based on asymmetrical matrices of word frequencies, where co-occurring words are used as variables and the documents as cases. These matrices were imported into UCINET and the visualizations were made with Pajek (for information on the methods, see Leydesdorff, 2004; Leydesdorff & Hellsten, 2005).

Table 1 first summarizes the numbers of all the downloads as contrasted to the reported numbers of results. It also summarizes the numbers of the title words in the downloaded results and the numbers of co-occurring title words, as well as the numbers of the co-occurring title words included in the analysis.

| | AV Jan 2004 | AV Feb 2004 | AV Apr 2004 | Ggl Jan 2004 | Ggl Feb 2004 | Ggl Apr 2004 |
|---|---|---|---|---|---|---|
| Nr. of reported records | 8222 | 8733 | 1239 | 5184 | 4955 | 3553 |
| Nr of retrieved records | 2106 | 2035 | 620 | 3068 | 2912 | 3115 |
| % downloads | 25.6 | 23.3 | 50.0 | 59.2 | 58.8 | 87.7 |
| Unique title words | 3495 | 3397 | 1821 | 4561 | 4328 | 4289 |
| Nr. of word occurrences | 9332 | 9004 | 3616 | 14597 | 14353 | 15020 |
| Threshold used | > 12 | > 12 | > 6 | > 18 | > 18 | > 18 |
| Words included in the semantic maps | 97 | 95 | 74 | 98 | 103 | 111 |
| Cosine ≥ 0.2 | 46 | 49 | 57 | 47 | 57 | 67 |

**Table 1** Summary of the downloads

Note the discrepancies between the reported numbers and the actually downloaded numbers in the first two rows of Table 1. These differences are the more striking since we used the 'site collapse' filter of the *AltaVista Advanced Search Engine* in order to exclude identical pages from the reported numbers. *AltaVista* provided higher reported numbers in January and February than *Google*, yet *Google* had higher numbers of actually downloadable records. It is intriguing that the actually downloaded numbers decrease much more slowly than the reported numbers. In the case of *Google* in April 2004 this number even rises, resulting in an exceptionally high percentage of downloaded records. The number of retrievable records seems to remain approximately the same, but the reported number may become more precise over time.

The rank order in the frequencies of words in the titles of Web pages are an indication of the similarity of structure in the Web search on 'frankenfoods' time-stamped for the year 2003. This information is summarized as rank-order correlations in Table 2.

| | | AV Jan | AV Feb | AV Apr | Ggl Jan | Ggl Feb | Ggl Apr4 |
|---|---|---|---|---|---|---|---|
| AV Jan | | 1.000 | .840(**) | .460(**) | .525(**) | .524(**) | .530(**) |
| | N | 3351 | 2965 | 959 | 1918 | 1806 | 1720 |
| AV Feb | | .840(**) | 1.000 | .495(**) | .564(**) | .556(**) | .541(**) |
| | N | 2965 | 3247 | 958 | 1902 | 1775 | 1689 |
| AV Apr | | .460(**) | .495(**) | 1.000 | .531(**) | .533(**) | .519(**) |
| | N | 959 | 958 | 1766 | 1168 | 1195 | 1172 |
| Ggl Jan | | .525(**) | .564(**) | .531(**) | 1.000 | .696(**) | .623(**) |
| | N | 1918 | 1902 | 1168 | 4360 | 2978 | 2537 |
| Ggl Feb | | .524(**) | .556(**) | .533(**) | .696(**) | 1.000 | .815(**) |
| | N | 1806 | 1775 | 1195 | 2978 | 4145 | 3448 |
| Ggl Apr | | .530(**) | .541(**) | .519(**) | .623(**) | .815(**) | 1.000 |
| | N | 1720 | 1689 | 1172 | 2537 | 3448 | 4077 |

** Correlation is significant at the 0.01 level (2-tailed).

**Table 2:** Rank correlation (Spearman's ρ) among word occurrences in different searches for the documents from the year 2003.

The word frequency distributions are most strongly correlated within each search engine, with the exception of *AltaVista* in April 2004. This auto-correlation reflects that the search engines are more likely to update URLs already included in their databases than to add new entries. Furthermore, the overlap between the search engines is expected to be moderate because of the different algorithms used for the crawling. The correlation coefficients prove that the change in *AltaVista* in April 2004 has fundamentally changed the operation of this search engine. Because of this structural change, we will only show the semantic maps for *AltaVista* as collected in January and February 2004. The correlations between *AltaVista* and *Google* are lower than those between the structures generated by each search engine, but their significance cannot be ignored.

What clusters can be discerned in these two related structures? The next two figures show the structure of the title words that co-occurred more than twelve times in the search results retrieved using *AltaVista*. The semantic maps are based on the searches conducted with *AltaVista* in January and February 2004 and represent the results attributed by the date stamps to the year 2003 (Figures 5 and 6).

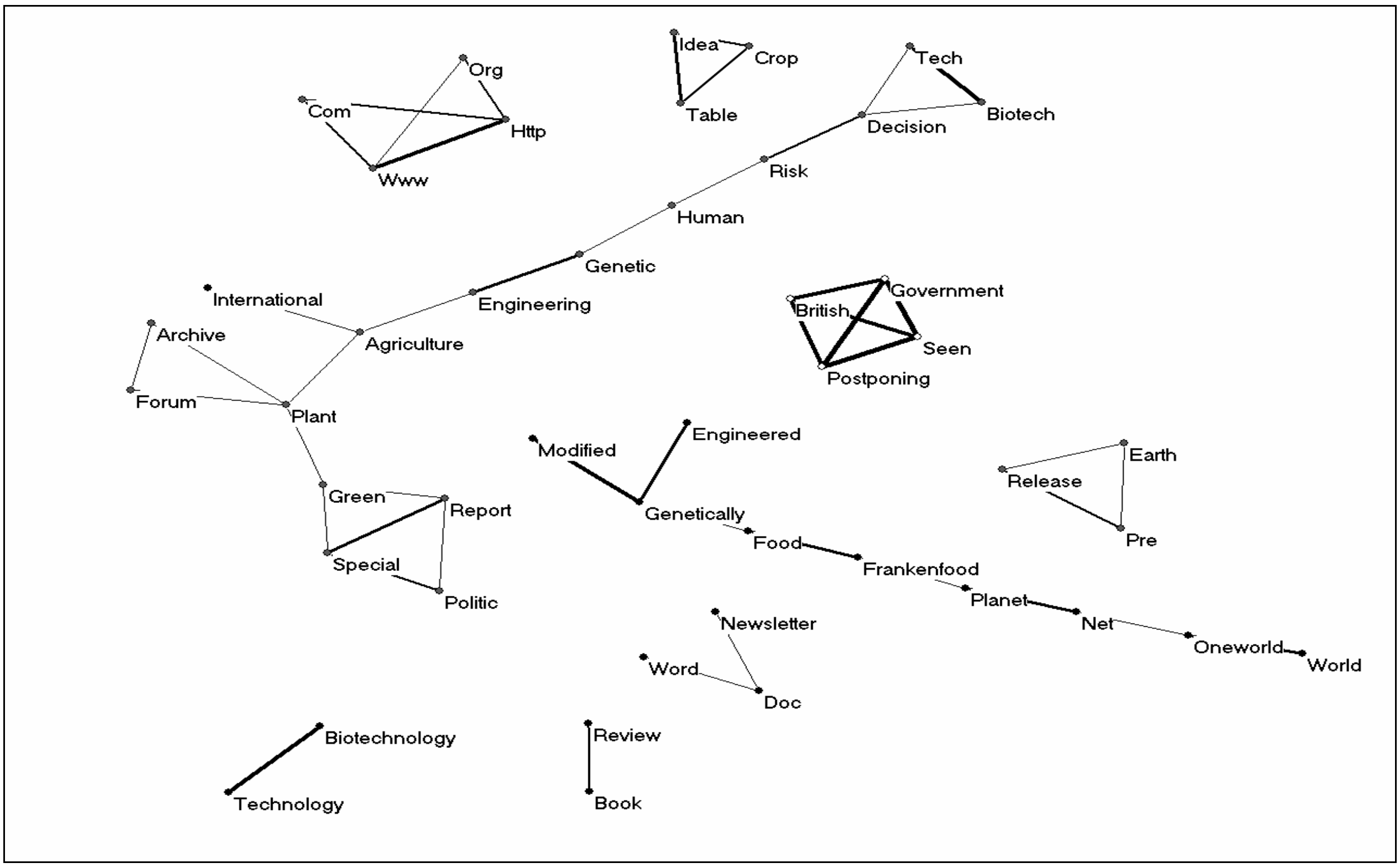

**Figure 5:** 46 words related at the level of cosine ≥ 0.2 and occurring more than 12 times in the 2106 records collected with the *AltaVista Advance Search Engine* in January 2004.

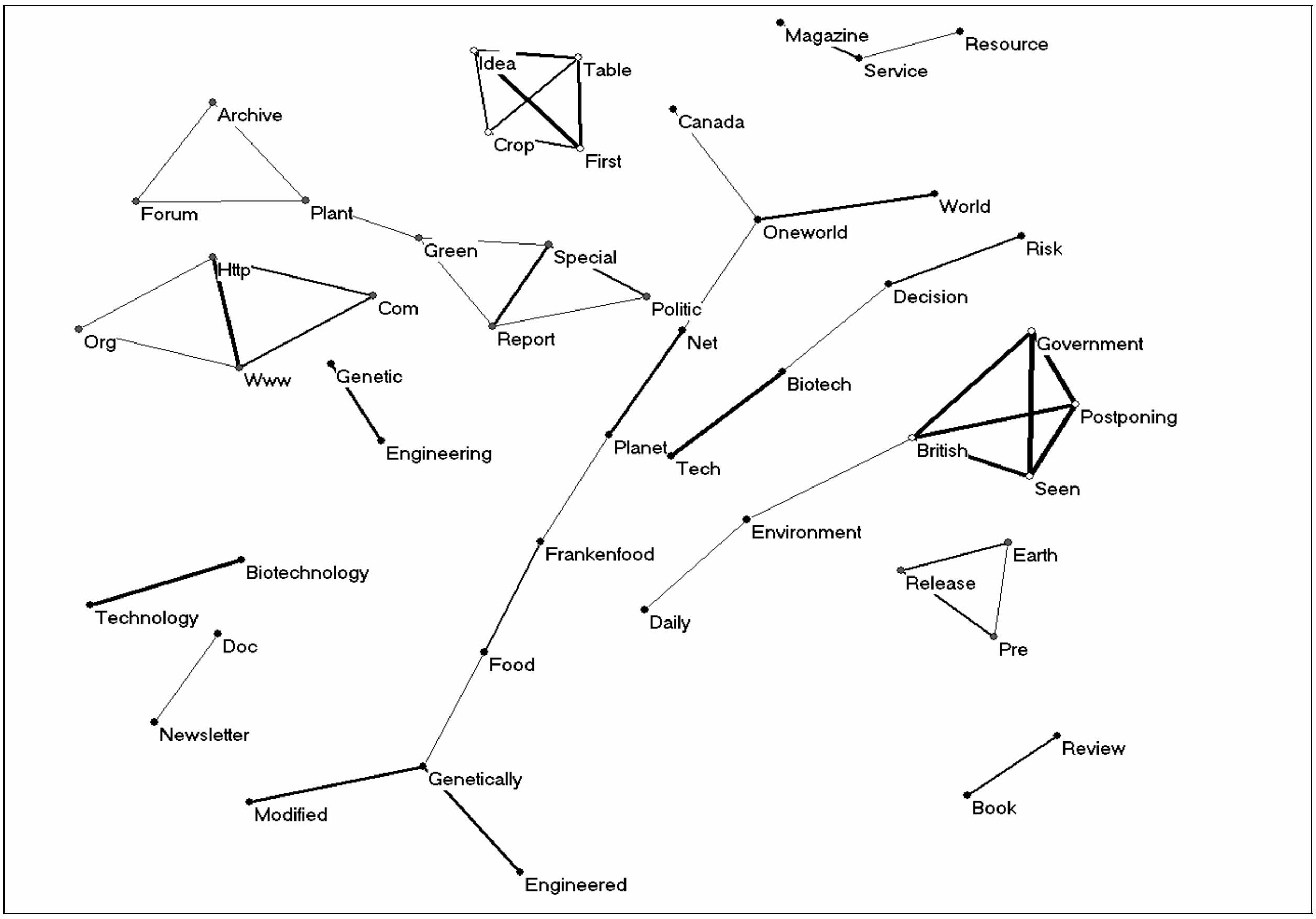

**Figure 6:** 49 words related at the level of cosine ≥ 0.2 and occurring more than 12 times in the 2035 records collected with the *AltaVista Advance Search Engine* in February 2004.

The picture of January is more informative in the sense of representing more connected—that is, larger—clusters than the one from February. Thus, the structure of the information erodes in the data over time. Fewer words explain more structure in data collected in January than in February. For example, the number of unrelated clusters increases from eight in January to twelve in February. At the same time, the results indicate continuity in the data. For example, the word clusters around Earth Press Release and Plant Forum Archive in both January and February are similar. Is this erosion also visible in the structures generated by *Google* (Figures 7 to 9)? The structure of the co-occurring title words in the set retrieved by *Google* clearly differs from that of *AltaVista* by representing unique word clusters around 'public debate' and 'new global headlines.' However, some clusters reflect the same words, such as Plant Forum Archive (Figure 7). The Web pages in this archive are stable.

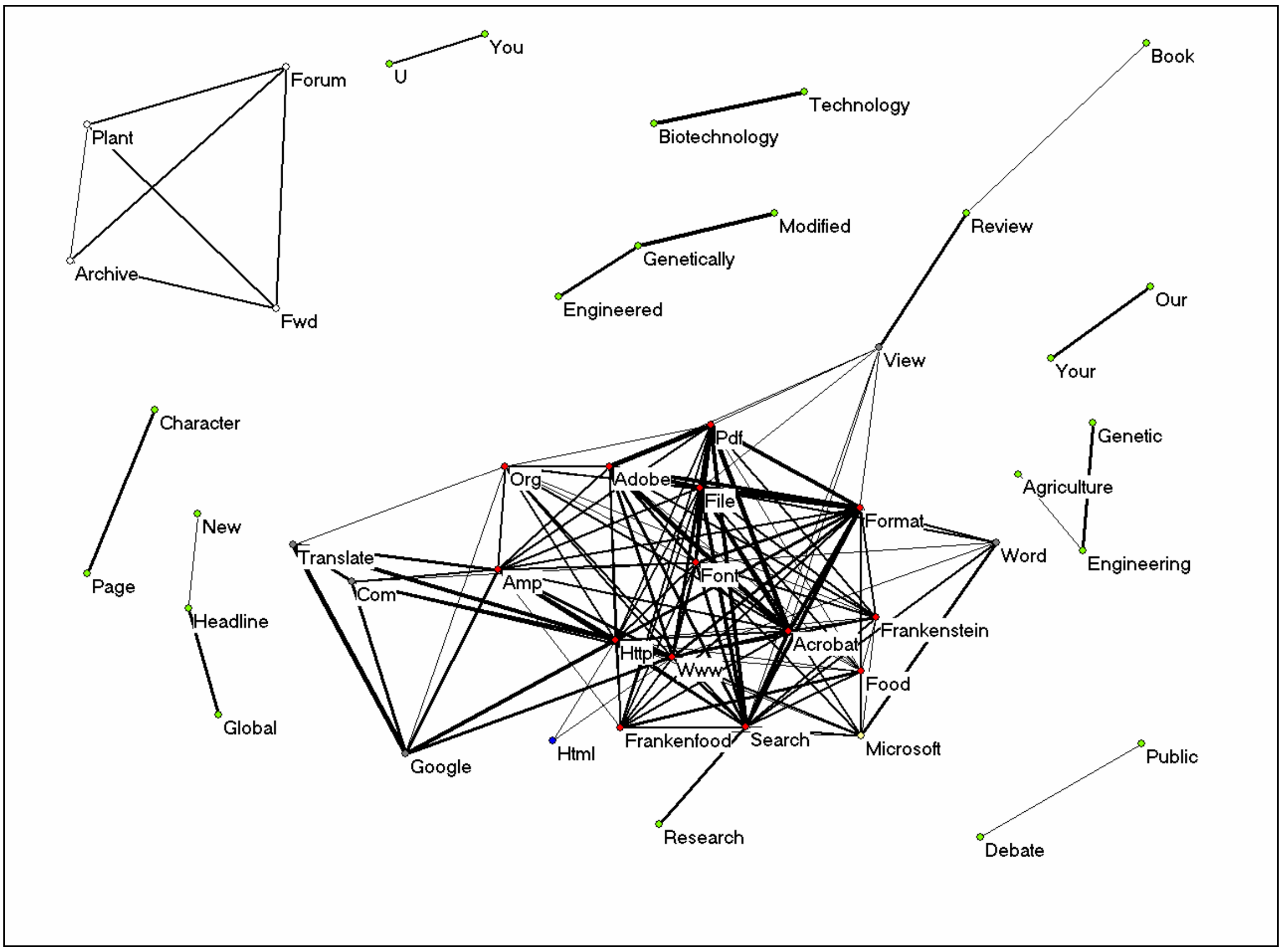

**Figure 7:** 47 words related at the level of cosine ≥ 0.2 and occurring more than 18 times in the 3068 records collected with the *Google Search Engine* in January 2004.

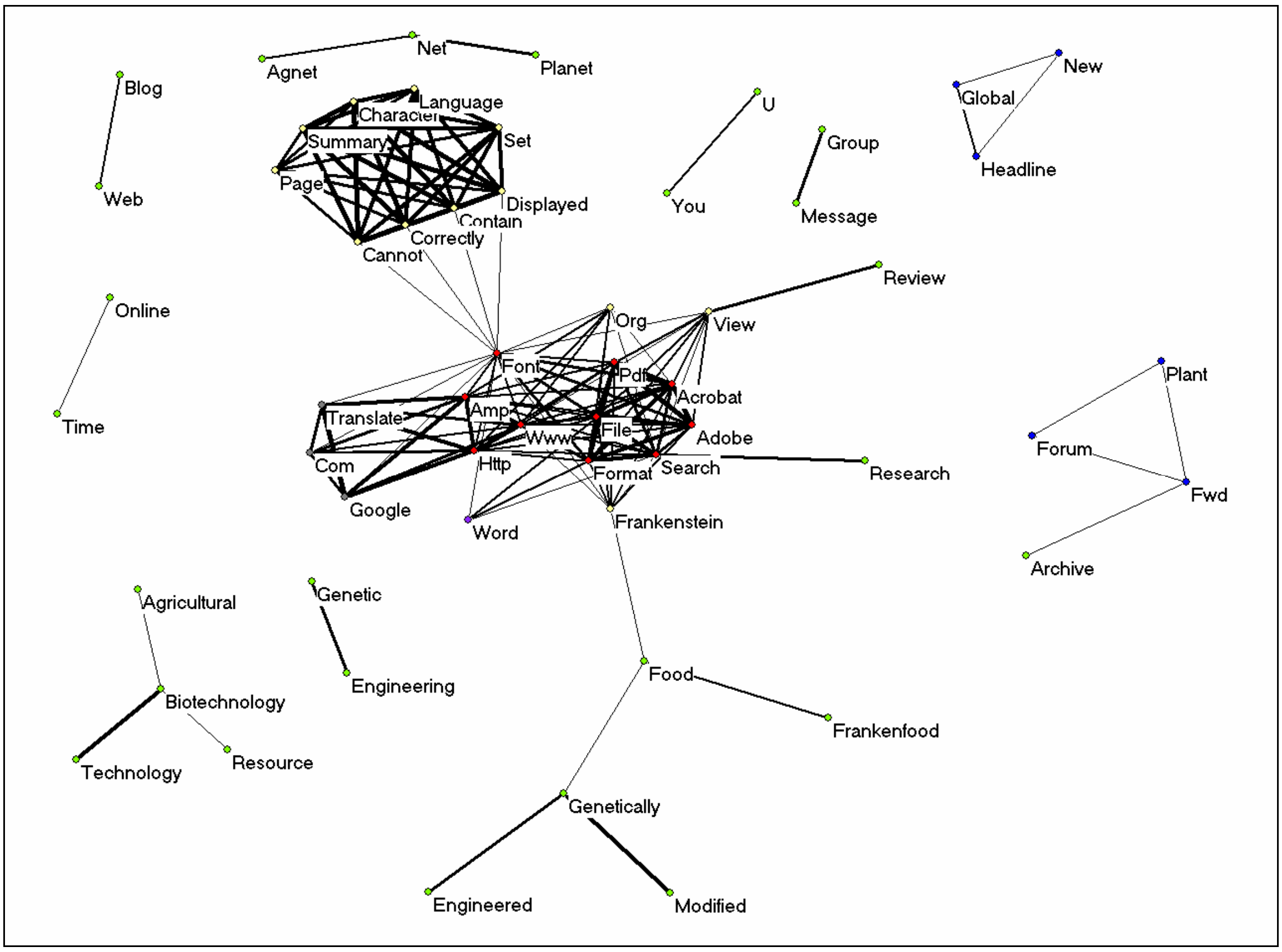

**Figure 8:** 57 words related at the level of cosine ≥ 0.2 and occurring more than 18 times in the 2912 records collected with the *Google Search Engine* in February 2004.

Between January and February, *Google* shows a change in terms of this analysis more than *AltaVista* did. Figure 8 exhibits the emergence of new word clusters, such as 'characters that cannot be correctly displayed,' 'time online,' 'Web blog,' and 'group message' that were not part of the title words collected in January. How the structure of the results further develops is presented in Figure 9.

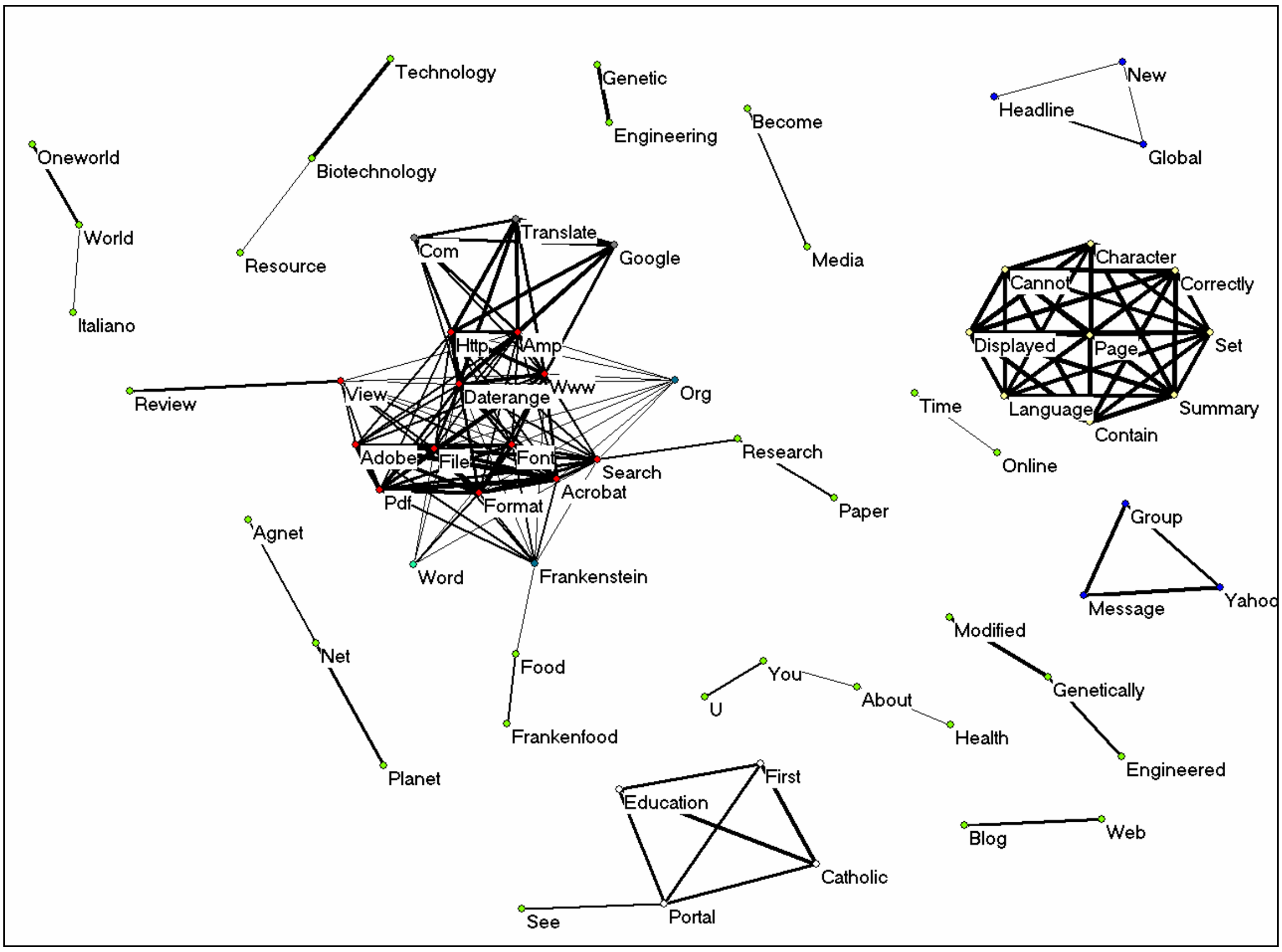

**Figure 9:** 67 words related at the level of cosine ≥ 0.2 and occurring more than 18 times in the 3115 records collected with the *Google Search Engine* in April 2004.

Figure 9 shows the further erosion of the structure of the co-occurring title words, and the emergence of one new cluster around 'education portal.' The number of disconnected clusters in the map rises from ten in January to fourteen in April.

In summary, the picture is a bit more complex than in the case of *AltaVista*, mainly because of the emergence of a cluster of words in February that is related to the message 'contains characters that cannot be correctly displayed.' Nevertheless, the trend for both *AltaVista* and *Google* is that of a loss of structure.

The maps based on the *Google* data appear more 'noisy' than those made from the January and February *AltaVista* data. This may be caused by the fact that the search engine of *Google* is based on using hyperlinks, whereas *AltaVista* uses keyword searches. A second difference between *Google* and *AltaVista* is shown in the cluster about non-displayable (i.e., non-latin) characters in the titles of the results. The algorithm of the search engine may have changed. We tentatively infer from this that *Google* has a wider window on non-English languages, precisely because it is not based on (English) keyword searches. Another explanation for this might be that *Google* indexes a wider variety of file types than most other search engines, for example, including pictures in the database (Reddy & Wouters, 2003). Furthermore, one should keep in mind that the search engines may crawl other parts of theWeb

## *Discussion*

We had three hypotheses about how the present of the search engines evolves over time. First, we expected that the distribution of the reported number of hits over the years 1995-2003 would show a strong bias in favour of the most recent year (relative to the date of the data downloading). Secondly, if it is true that Web sites are constantly overwritten with newer date stamps, then we expected also to find a decrease in the total number of hits for the months before the most recent one (again relative to the date of the measurement). The results confirm both these hypotheses. We expected substantial erosion over time in the sense that search engines not only re-write the past but also forget the past. This dynamics of reconstruction from the perspective of hindsight is extremely relevant because the reorganization feeds back on the overall growth of the Web. Thirdly, we expected that the structure of the information erodes over time, and our results confirmed this. The past in the internet is constantly overwritten from a hindsight perspective that affects the numbers of the results as well as the actual Web pages the search engines retrieve. Hence, the presents from where the data is collected affect the search results considerably.

Our third finding—that both search engines not only lose information quantitatively, but that they also erase the structures entailed in the relationships between words of Web page titles—may be even more important from a social science perspective than the updating of the time stamps of Web pages and Web documents as such. What is particularly striking is that we are not dealing here primarily with instabilities. On the contrary, in many ways the updating mechanism of search engines is remarkably stable and systematic.

This does not mean we did not meet any instabilities. We experienced two types of instabilities: first, the fundamental restructuring of *AltaVista* in April 2004, which made its results before and after the re-organisation unrelated; and second, the variable difference between the reported number of records and the number of records that could actually be retrieved.

However, the main phenomenon we have dealt with in our experiments is not instability, but the systematic erasure of both the historical record and the structures in informational and semantic networks. This is caused by the fact that the search engines are tied to the updating cycles of the Web and the internet, rather than to the historical development of its structure. The structure in the information on the Web at any given moment of time is the result of relations that exist at the moment of the creation of Web pages and the various updating mechanisms that we have shown in our experiments. Although we do not know how these two forces balance each other during the period of incorporating the Web pages into the search engines, the lag times seem different between the two search engines under study. However, we have shown that over time, the structure of information as represented in the relationships between words is determined by the updating frequencies, and as a consequence, erosion of this structure is taking place.

In other words, the fact that the search engines of the internet and the Web are actually a complex network of networks, each with its own updating cycles, leads to a loss of structure in the word clusters. This raises the question of whether this loss of structure may have a finite window: are networks of information after a longer period of time becoming more stable, or does the erosion of structure continue? We would expect the latter to prevail because of the continuation of the relevant operations. Another relevant question would be whether particular configurations of networks ‘travel’ from year to year. It would also be interesting to know to what extent the specifics of the search strings (‘frankenfoods’ in our case) influence the types of networks generated and their decay times.

## *Conclusion*

We have shown that the search engines *AltaVista* and *Google* systematically relocate the time stamps of Web documents in their databases from the more distant past into the present and the very recent past. Second, the search engines delete documents from the year to which they were initially assigned. This leads to a loss of information in the historical record on the Web as represented in the search engine databases. Third, information is lost not only in the quantitative sense of documents disappearing from the historical record, but also in the sense of a loss of structure in the semantic networks.

This does not mean that search engines are completely useless for scholarly research or do not represent a significant phenomenon on the Web. On the contrary, our results confirm that we can appreciate search engines as the clocks of the internet, ticking at different frequencies and possibly leading to multiple presents. They provide representations that indicate the updating frequencies of both the Web and the underlying internet. How should we interpret this? We are dealing with complex interactions of the updating frequencies of individual Web pages by their authors or Web masters; the updating frequencies and mechanisms of the structures in which these Web pages are positioned; the frequencies with which these Web pages are being visited by search engine crawlers; the extent to which 'old' Web pages are retained in search engine databases although more recent version of that 'same' Web page have been added; the overall growth of the Web; and, last but not least, by the updating frequencies of the sorting algorithm of the search engine and its presentation mechanisms. All these frequencies can be expected to differ. Moreover, each search will be influenced in different ways by the various frequencies. Each search engine can therefore be said to represent not one updating frequency but a frequency distribution or spectrum of frequencies (including very slow changes for static Web pages). This spectrum may be specific to a given search engine in a particular period of its existence (Smolensky, 1986).

What does this mean in relation to time and temporality? As clocks of the internet, search engines realize the present as a collection of extended presents that can exist in parallel on the Web. In other words, time is being represented as realities that coexist in space. The concept of the 'extended present' (Nowotny, 1994) has been used mainly to indicate the dominance of the present over the past and future, and to broaden the concept of the present from a fleeting point in time to a spectrum of actualities. We propose to add the notion of fragmentation to this concept. We are not dealing with one extended present, but with a multiplicity of partly conflicting presents. The frequency with which these extended presents are being updated in turn does not have to relate to the development of the actors whose presence is here represented. This differentiation may contribute to the feeling of being overwhelmed by the information overload of the system (Luhmann, 1996).

We have interpreted search engines as clocks of the internet driven by the interaction of a variety of updating frequencies. We have shown with our experiments that these clocks not only run at different frequencies depending on the 'present' of the searches and the search engine in question, but also reconstruct the pasts in very different terms. Each search engine differs in the combination of these frequencies and their selection, resulting in different lag times and information restructuring windows. The question of how we can make better use of search engines in scholarly research to unveil the overall updating cycles that dominate the Web and particular domains of the Web becomes an interesting research question that should be put on the agenda. Search engines are the 'clocks' of the Web, but rather strange ones that act more like the clocks of Salvador Dali than those of Christiaan Huygens.